\documentclass[12pt]{article}
\usepackage{times}
\usepackage{multicol}
\usepackage[sort]{natbib}
\usepackage{epsfig}
\usepackage{url}

\def\apjl{ApJL}
\def\apjs{ApJS}

\def\aap{{\em A.\&A}}

\usepackage[layout=modern]{advancedcoverpage}

\title{The Impact of the Astro2010 Recommendations on Variable Star Science}

\author{\emph{\underline{Corresponding Authors}}\\
 \vskip .1in
  {\bf Lucianne M. Walkowicz}\\
  Department of Astronomy, University of California Berkeley\\
  {\tt lucianne@astro.berkeley.edu}\\ 
  phone: (510) 642--6931\\
 \vskip .05in
 {\bf Andrew C. Becker}\\
 Department of Astronomy, University of Washington\\
 {\tt becker@astro.washington.edu}\\
 phone: (206) 685--0542\\
\vskip .05in
{\em{\underline{Authors}}}\\
\vskip .05in
  Scott F. Anderson, Department of Astronomy, University of Washington\\
  Joshua S. Bloom, Department of Astronomy, University of California Berkeley\\
  Leonid Georgiev, Universidad Autonoma de Mexico\\
  Josh Grindlay, Harvard--Smithsonian Center for Astrophysics\\
  Steve Howell, National Optical Astronomy Observatory\\
  Knox Long, Space Telescope Science Institute\\
  Anjum Mukadam, Department of Astronomy, University of Washington\\
  Andrej Pr\v sa, Villanova University\\
  Joshua Pepper, Villanova University\\
  Arne Rau, California Institute of Technology\\
  Branimir Sesar, Department of Astronomy, University of Washington\\
  Nicole Silvestri, Department of Astronomy, University of Washington\\
  Nathan Smith, Department of Astronomy, University of California Berkeley\\
  Keivan Stassun, Vanderbilt University\\
  Paula Szkody, Department of Astronomy, University of Washington\\
}

\presentedto{\underline{Science Frontier Panels:} \\
{\it Stars and Stellar Evolution (SSE)}}

\begin{document}
\maketitle

\centerline{\bf \Large Abstract}
\smallskip

  The next decade of survey astronomy has the potential to transform
  our knowledge of variable stars.  Stellar variability
  underpins our knowledge of the
  cosmological distance ladder, and provides direct tests of stellar
  formation and evolution theory.  Variable stars can also be
  used to probe the fundamental physics of gravity and degenerate
  material in ways that are otherwise impossible in the laboratory.
  The computational and engineering advances of the past decade have
  made large--scale, time--domain surveys an immediate reality.  Some
  surveys proposed for the next decade promise to gather more data
  than in the prior cumulative history of astronomy.  The actual
  implementation of these surveys will have broad implications for the
  types of science that will be enabled.  We examine the design
  considerations for an optimal time--domain photometric survey
  dedicated to variable star science, including : observing cadence,
  wavelength coverage, photometric and astrometric accuracy,
  single--epoch and cumulative depth, overall sky coverage, and data
  access by the broader astronomical community.  The best surveys must
  combine aspects from each of these considerations to fully realize
  the potential for the next decade of time--domain science.

\section{Introduction}
%


%
%

Recent surveys such as MACHO, OGLE-II and III, HAT, ASAS, SuperWASP,
HIPPARCOS, and others have provided insight into the distribution of
stellar variables as tracers of local structure, as well as the
intrinsic nature of variability in the objects themselves.  
In the coming decade, the 
%
next generation of large--scale surveys will build on these
discoveries as well as raise new questions.
We argue that an optical, time--domain survey lies at the critical
intersection of feasibility, science return, and discovery potential
if the data can be shared with the community and its vast diversity of
domain experts.
All--sky (or all--available--sky) surveys allow us to test in an
unbiased manner the power--spectrum distribution of matter in the
Galaxy and in the Universe.  
Through the production of very large samples of known variable types,
these surveys will enable statistical studies of the ensemble to
definitively characterize the class boundaries and find outliers from
the mean \citep{2007AJ....134.2398C}.  
These surveys will also lead to the discovery of theoretically
predicted populations that have not yet been observed
\citep[eg. pulsating brown dwarfs;][]{2005A&A...432L..57P} as well as
systems found in unique configurations \citep[eg. eclipsing AM Canum
Venaticorum systems;][]{2005AJ....130.2230A}.
A prime example of this paradigm is the Sloan Digital Sky Survey
(SDSS), a single--epoch photometric survey of more than $10^4$ square
degrees determined to have the highest recent impact in astronomy
based on a citation analysis of papers by \cite{2009arXiv0901.4552M}.
A time--domain survey with immediate public data access is the next
logical step in the evolution of astrophysical surveys, and will have
a transformative impact across all disciplines of astronomy.
%
%
For a full overview of recent results from stellar variability
surveys, we refer the reader to the review by Eyer \& Mowlavi (2008).
In this white paper, we highlight 
%
high--impact projects that will be made possible by large--scale
time--resolved samples of known variable classes.

\subsection{Calibrating Cosmic Distance: Cepheids, RR Lyrae and Miras}
Pulsating stars are the preeminent distance indicators, both as local rungs in the cosmic
distance ladder and as tomographic tools to study the 3-D structure of
the Galaxy.  With RR Lyrae and Cepheid
variables, uncertainties in distances measurements coupled with
unknown dust and metallicity corrections translate directly into
uncertainties in cosmological parameters
\citep{2003LNP...635...85B,2008ApJ...679...52T}.  
%
Advances in image subtraction techniques applied to repeated deep
imaging have extended the range to which these stars may be detected
in external galaxies.  Ensembles of these stars derived from a range
of star formation histories will enable solutions for the metallicity
dependency of their period--luminosity relationships, making them more
precise distance indicators.  The superior depth provided by new endeavors like LSST
(\url{http://www.lsst.org}) will make statistically significant samples of extragalactic variables a reality.

Beyond the optical, equivalently deep
infrared observations (e.g. SASIR; \url{http://www.sasir.org}) would
allow for a precise calibration of the period--luminosity relations at
infrared wavelengths, skirting issues of dust extinction (and in
combination with optical observations allow the construction of a
wide--field dust map of unprecedented spatial precision). In turn,
AO-enabled 30 meter telescopes would then be able to measure precise
distances well-beyond the local group and finally fix the rungs of the
cosmic distance ladder out to $\sim 25$\,Mpc.

These new observations will also improve our understanding of the structure and history of our own Galaxy: RR Lyrae are essential tracers of structure and metallicity in the
Milky Way \citep{1995rls..book.....S}.  SDSS detected RR Lyrae to
distances of $\sim$100 kpc \citep{2007AJ....134.2236S}, uncovering
halo substructure through clumps in the RR Lyrae spatial distribution.
LSST will detect RR Lyrae to distances of
$\sim$400 kpc, providing extensive tests of hierarchical galaxy
formation models \citep[e.g.][]{2001ApJ...548...33B}.

Mira variables are the most luminous of potential distance indicators,
but their use has been observationally impractical due to their
extremely long periods (hundreds of days).  However, their
luminosities give the Mira population great potential as a cosmic
distance indicator, bridging the distances where vast amounts of
stellar variables are found, and where vast numbers of Type Ia
Supernova are found.  
%
%
%
%
{\bf Finally, we mention that the ability to detect variables in
external galaxies opens the window to the detections of
extraordinarily rare, one--in--a--galaxy, phenomena.}

\subsection{Luminous Blue Variables and Cool Supergiants}

The scarcity of high mass stars poses a serious challenge to our
understanding of stellar evolution atop the HR diagram.  As O--type
stars evolve off the Main Sequence their violent death throes can be
characterized by extreme mass loss and explosive outbursts, which are
short--lived and possibly intermittent.  There are only a handful of
nearby massive stars that are caught in this phase at any given time
(as in the case of $\eta$ Car), which makes it very difficult to
connect distant explosions (supernovae and GRBs) to their underlying
stellar populations.


The improved depth and breadth of proposed surveys will make these
extremely luminous stars accessible in other galaxies, allowing the
community to continue the expand on the early work of
\cite{1953ApJ...118..353H} and \cite{1968ApJ...151..825T}.
Time--resolved observations of variability in this new sample will
quantify the statistical distribution of time--dependent mass--loss
rate, luminosity, radiated energy, total mass ejected, duration of
outbursts, time between outbursts, and connections to the
pre--outburst stars.
New observations will inform models of massive star evolution,
including its dependence on metallicity, providing prescriptions for
the time--dependent properties mentioned above so that they can be
included in stellar evolution codes in a meaningful way.  In a
complementary fashion, further study of these stars will also improve
our understanding of galactic feedback and enrichment of the
interstellar medium.

Another open question regards the true nature of core collapse
supernova (CCSN) progenitors. A large sample of evolved massive stars
will propel our understanding of the diversity of CCSN progenitors.
{\bf Among the large sample of luminous stars monitored in nearby
galaxies, some may explode while they are being monitored}. This will
provide not only an estimate of the star's pre--explosion luminosity
and temperature, but also its variability and potential instability in
the final years of its life.
It is also possible that stellar population studies of the surrounding
field stars can constrain the local star formation history, and thus
constrain the delay time between star formation and core collapse.



\subsection{Galactic Chronometers: Pulsating White Dwarfs} 
\label{sec:wd}

Isolated white dwarf variables are found in four distinct instability
strips located in different temperature regimes.  All of the white
dwarf variables exhibit nonradial gravity--mode pulsations, and most of
these pulsators are multi--mode, showing at least a few modes at the
same time.  Differences in pulsation amongst hydrogen--atmosphere white dwarfs are predominantly affected by differences in temperature, as these stars are otherwise very
homogeneous.  As white dwarfs are simply
radiating away leftover thermal energy, the luminosity of a given white dwarf is a straightforward function of age.  A large sample of white
dwarfs will thus allow calibration of the white dwarf cooling curves.

The observed pulsational periods are tens to thousands of seconds,
with amplitudes as large as 10\% and diminishing in amplitude as the
stars reach the edges of their instability strip.  There are four
known white dwarf instability strips, although their boundaries are
not well defined.  These come from the Hydrogen atmosphere (DA),
Helium atmosphere (DB), hot (DO), and recently discovered Carbon
atmosphere \citep{2009arXiv0901.3489F} white dwarfs pulsators.
A survey observing these stars once a night may not entirely resolve
the pulsational period.  However, repeat measurements will uncover
those stars that exhibit a scatter in their lightcurves larger than
the measurement uncertainties.  These ensembles of stars can then be
used to experimentally {\it define} the extent of the white dwarf
instability sequences.
A similar illustrative experiment was carried out by
\cite{2007AJ....134..973I}, using data from SDSS repeat observations.
Figure~\ref{fig:wd} shows the colors of non--variable ($\sigma_{g} <
0.05; \sigma_{r} < 0.05$) objects near the white dwarf cooling
sequences.  The rightmost panel shows single--epoch colors taken from
SDSS DR5.  The left panel shows the averaged colors of the objects
over $\sim 10$ epochs.  Multiple sequences are apparent which were not
evident in the single--epoch photometry, two of which correspond to
the cooling curves of H and He white dwarfs
\citep{1995ApJ...443..764B}.
{\bf These are fundamental tests of degenerate matter that cannot be
replicated in the lab.}  This includes constraints on exotic particles
such as axions and plasmon neutrinos \citep{2007PhDT........13K}.

{
\begin{figure*}[t]
\begin{minipage}{0.40\textwidth}
  \caption{\footnotesize SDSS color--color diagram for objects near
    the white dwarf cooling sequence.  The {\it right} panel shows the
    colors for all sources seen to be non--variable over many epochs,
    but only shows their photometric measurements from one epoch.  The
    {\it left} panel shows the mean colors for these objects over all
    epochs, and resolves cooling sequences much less apparent in the
    single epoch photometry.  Adapted from Figure~24 of
    \cite{2007AJ....134..973I}.  }
\medskip \hrule \normalsize
\label{fig:wd}
\end{minipage}
\begin{minipage}{0.60\textwidth}
\centerline{\psfig{file=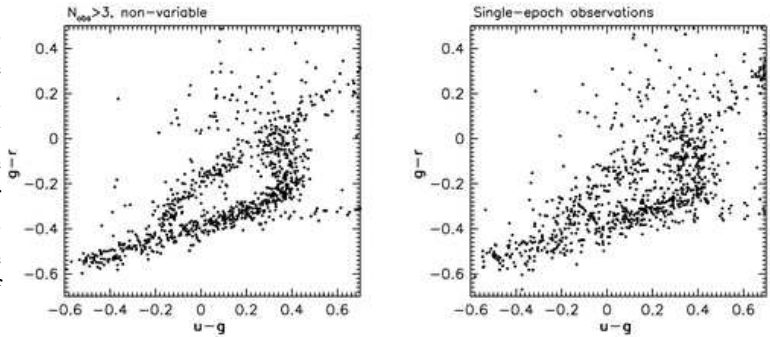,width=\textwidth}}
\end{minipage}
\end{figure*}
}

\subsection{Close and Interacting Binaries}

New surveys will also allow for the first robust, volume--limited
population census of stellar remnants within our Galaxy.  Of these
remnants, the most numerous and easiest to study observationally are
the interacting white dwarf binaries.  These systems include X-ray
binaries and symbiotic stars (for more massive stars), cataclysmic
variables (CVs -- including novae, dwarf novae and novalikes) and
ultimately, the double--degenerate AM CVn systems
(Tutukov \& Yungelson 1996) and Ultra-compact X-ray binaries
(Nelemans et al. 2006).

As a large fraction of stars are binaries, it is important to
understand the effects of binary membership on stellar evolution.
Binary interaction also alters the stellar evolution process in many
ways that can result in spectacular transient and variable phenomena.
Faint, wide surveys will find binaries at greater distances, with
lower mass companions and lower mass transfer rates than previously
possible.
A time--domain study of these systems will allow us to characterize
the patterns of accretion across the spectrum of $\dot{M}$ in a
systematic fashion.
%
%
The majority of CVs in the Galaxy are predicted to have low
mass--transfer rates and thus be intrinsically faint \citep[$22 < V <
25$;][]{1998ASPC..137..207P,2001ApJ...550..897H} just below the
limiting magnitude of recent all--sky surveys such as SDSS but
distinguishable by both their colors and intrinsic variability in
next--generation surveys.
Detection of this theoretically predicted population will provide
fundamental tests of many aspects of stellar evolution -- how stars
form into giants and super--giants, how mass loss and core evolution
proceeds as the stars evolve into white dwarfs, and the process of
angular momentum loss -- ending with the majority population of white
dwarf binaries.
%
%
{\bf Direct detection of CVs across the spectrum of mass transfer
rate, especially detecting the faint majority population, will help us
to understand the diversity seen in cosmological Type Ia supernovae
\citep{2005ASSL..332...97F} by observing the diversity in their
progenitor Galactic systems. }
%
Finally, the space density and orbital period distribution of detected
AM CVn are key ingredients in understanding the gravitational wave sky
expected to be unveiled in the next decade
\citep{2004MNRAS.349..181N}.

\subsection{Additional Science Topics}

We enumerate other unique science that is enabled by a broad
time--domain survey, and that may be emphasized more thoroughly in
other white papers.  This includes :

\begin{itemize}

\item An unbiased measurement of the period--amplitude diagram of
  stellar variability;

\item The frequency and amplitude of flares on M--dwarf stars, which
  affects their astrobiological viability as hosts for habitable worlds; 

\item The measurement of the masses of isolated objects using
  gravitational microlensing.  This technique is sensitive to
  planetary--mass to black hole--mass objects, and is one channel to
  finding exotic types of compact objects;

\item Understanding the diversity of cosmological transients by
  characterizing the foreground of Galactic phenomena.

\end{itemize}

Finally, we emphasize the strength of time--domain surveys : {\bf
repeat observations enable both the identification of periodic, non-periodic, and new classes of variable source as well as more accurate properties for non-variable sources.}

\section{Technical Feasibility : Survey Design Considerations}


In the following section, we discuss the design considerations for an
optimal time--domain photometric survey dedicated to variable star
science, including observing cadence, wavelength coverage, photometric
and astrometric accuracy, and data access by the broader astronomical
community.

\subsection{Photometric Accuracy}

The internal photometric calibration of a survey reflects the degree
to which it can detect variability.  
%
%
The newest frontiers are enabled by survey--scale photometry
repeatable at the $1\%$ level or below, which enables novel science
such as estimating the ages of stars through their rotation period
(where the variability comes from star spots) or detecting solar--type
oscillations on other stars \citep{1993AJ....106.2441G}.
The limits on internal calibration come from a combination of
irreducible random noise, and systematic effects that need to be
minimized in the experimental design.  
Recent optical wide--field ground--based surveys have achieved the
benchmark of $\sim 1\%$ repeatability in photometric
measurements \citep{2007AJ....134..973I}.
The fundamental limitations to this number come from uncertain
calibration of the instruments and from the uncertain knowledge of the
instantaneous transmission profile of the atmosphere, which causes
variations in the effective filter of the system.  
%
%
Space--based observations avoid many of these issues, and with careful
metrology can reach the photon noise limit.

All known pulsating variables with the exception of white dwarfs have
M$_V$$<$+2.5, primarily due to observational selection effects in
areal coverage, limiting magnitude, observed sample size, photometric
precision, and time coverage.  An expansion of capability in each of
these areas will greatly increase the discovery space even for normal
pulsators.
\cite{Howell2008} discusses the relationship between the detectable
variable fraction and the photometric precision of a given survey,
citing an exponential increase in the fraction of observed variable
sources as a function of improved photometric precision.  For a survey
with internal calibration of 0.005 magnitudes, approximately 12\% of
all sources should show signs of variability.  For a survey with
internal calibration of 0.001 magnitudes, $\sim 50\%$ of objects
should show variability.  {\bf Internal photometric accuracy is
clearly a proxy for how much science can be gleaned from the data, and
should be optimized at high priority.}





\subsection{Astrometric Accuracy}


While not addressed in previous section, fundamental science can be
done using the {\it spatial} variability of stars.  The astrometric
accuracy of a survey impacts its ability to measure parallaxes and
proper motions.  Parallaxes in particular set the foundation for
almost all astronomical measurements, being the only direct distance
measure, allowing the measurement of intrinsic stellar luminosities,
and measuring three--dimensional spatial structure.
From the ground, atmospheric effects dominate astrometric accuracy for
the bright sources, setting a systematic floor of approximately 10
milli--arcseconds (mas) per visit on spatial scales of arcminutes.
From space, the PSF sets the limit of astrometric accuracy.
Repeat observations of objects will average over these uncorrelated
systematic effects and improve overall astrometric accuracy.
In particular, the precision of parallax measurements improves with
time (scaling as $t^{-1/2}$) by building up repeated observations of
the parallax signal.
Proper motion precision improves with time ($t^{-3/2}$) by allowing
the proper motion vector to grow in amplitude.
Thus long duration surveys are desirable in this regard.





\subsection{Sampling Window in Time}

The cadence at which a survey revisits a given object defines the
timescales at which the survey is sensitive to its variability.  For
short, periodic variability a general survey is unlikely to Nyquist
sample the full cycle.  However, under the assumption of stability,
sampling the signal over many cycles can lead to its period and shape
recovery with little difficulty aside from sampling aliases.  To
address this latter issue, a ``rolling cadence'' with revisits at many
timescales will help to remove sampling aliases, as well as to fully
explore the timescales at which variability may be found.

Variability at short timescales is a regime that has been undersampled
by modern surveys.  The vast phenomenology known to exist there, from
M--dwarf flares to optically bright Gamma Ray Bursts, have yet to be
addressed in an unbiased manner.  In the next decade surveys will
uncover numerous instances of these phenomena, $\sim$30 per day per square
degree for M--dwarf flares (Kowalski et al. 2009).  Of requisite need
is an initiative by the surveys to release their data to the community
on a timescale commensurate with, in fact shorter than, the
phenomenological timescale to trigger follow--up and study by the
community.

%
%

The overall length of the survey also impacts our ability to
characterize an object's variability.  This statement applies equally
well to variability on timescales longer than the survey's lifetime,
as well as to exploring variations within shorter timescale phenomena,
such as period evolution in eclipsing binary systems and in pulsating
variable stars.  Although space--based surveys may provide certain
other observational advantages, the harsh environs of space and the
difficulty of hardware maintenance ultimately limits their lifespan.



\subsection{Sampling Window in Wavelength}

Color information, when combined with variability amplitude and
timescale, is an essential ingredient in successful object
classification \citep{2007AJ....134.2236S}. The choice of wavelength
sampling, and in particular cadence in a given filter, strongly
affects the science return possible from a given survey.
%
%
Different wavelength ranges often trace fundamentally different
physical processes.  For example, cataclysmic variables that possess accretion
disks display their greatest variability in the blue, but magnetic CVs
(``polars'') are most variable in the red due to cyclotron emission.



Because of opacity in the near--ultraviolet, $u$--band observations
are powerful discriminators of metallicity.  In addition, many
transient phenomena (such as M dwarf flares) have their highest
contrast in the blue.
%
%
In any ground--based survey, however, $u$--band observations are
difficult due to scattering by the Earth's atmosphere.
%
At the other extreme, infrared observations are powerful means to
discover the vast numbers of our M, L, and T--dwarf neighbors, as
evidenced by 2MASS observations \citep{2008AJ....136.1290R}.
Additionally, the time--domain is essentially unexplored in the
infrared.  
%
%
Eclipsing M--dwarf systems should be found in abundance along with
other interesting, intrinsically red, systems like the young stellar
objects KH-15D \citep{2003ApJ...593L.121W}.


\subsection{Outreach and Data Access}
 
A final consideration concerns the public access to, and timely
release of, the survey data.
It is highly unlikely that the survey teams will be able to
exhaustively mine their data for variability science.
Surveys that commit to enabling data access by the entire astronomical
community therefore maximize the overall scientific gains.
For time--domain science in particular, anything less than nearly
real--time release of the data will blunt the potential science impact
of the observations.
%
%
%
Without real--time data reduction and dissemination, the recognition
and study of short timescale, transient phenomena is not likely to be
achieved.

A commitment to real--time and data--release community access will
leverage the considerable investment already made into standards to
communicate such information efficiently, including development
efforts of the Virtual Observatory and the VOEvent infrastructure.
Such an integration will enable science far beyond the members of the
survey teams.  The entire astronomical community will benefit, from
professional astronomers supervising the next--generation of
autonomous follow--up telescopes
\citep{2008AN....329..269H,2006ASPC..351..751B}
%
%
to amateur astronomers curious to take a peek at the newest transient
event.  Access to the same survey--quality data stream will both
broaden and strengthen the community.

On similar footing is the degree of Education and Public Outreach
proposed by each survey.  The educational opportunities available with
such large time--domain datasets cannot be overstated.  Surveys
committing to, for example, providing a feed to Sky in Google Earth
("Google Sky") are enabling K--12 science teachers to communicate
daily the concept of our varying Universe.  
%
%
The ideal survey will also be active in outreach by
participating in initiatives such as NSF's Partnerships in Astronomy
\& Astrophysics Research and Education (PAARE) and the DOE / NSF
Faculty and Student Teams (FaST) program.  We argue strongly that this
is an attribute that should not be overlooked when ranking proposed
programs.


\begin{multicols}{3}
\begin{scriptsize}
\bibliography{whitepaper_variables}
\end{scriptsize}
\end{multicols}

\end{document}